\documentstyle{article}
\begin{document}
\begin{center}
\Large\textbf{ Hyperfine splitting in noncommutative  spaces.}
\end{center}

\begin{center}
\textbf{S. A. Alavi}

\textit{Department of Physics,  Tarbiat Moallem university of
Sabzevar, Sabzevar, P. O. Box 397,
 Iran}\\

\textit{}\\

\textit{Email: alavi@sttu.ac.ir}
 \end{center}



\emph{ We study the hyperfine splitting in the framework of the
noncommutative quantum mechanics(NCQM) developed in the
literature. The results show deviations from the usual quantum
mechanics.  We show that
 the energy difference between  two excited $F=I+\frac{1}{2}$ and the ground   $F=I-\frac{1}{2}$ states in
 a noncommutative space(NCS) is bigger than the one in commutative case, so the radiation wavelength in $NCS_{s}$
  must be shorter than the radiation wavelength in commutative spaces. We also find an upper bound for
 the noncommutativity parameter.}\\

 \large\textbf{Introduction.}\\

 \large{ Recently there have been much interest in the study of physics in noncommutative spaces($NCS_{s}$), not only because
 the  NCS is necessary when one studies the low energy effective theory of D-brane with $B$ field background, but also
 because
 at the very tiny string scale or in the  very high energy situation, the effects of noncommutativity of space may appear.
 In the literature the noncommutative quantum mechanics(NCQM) and noncommutative quantum field theory have been studied
 extensively and the main approach is based on the Weyl-Moyal correspondence which amounts to replacing the usual
 product by the star product in a noncommutative space. In the usual quantum mechanics,
  the coordinates and momenta have the following commutation relations :
\begin{equation}
[x_{i},x_{j}]=0\hspace{1.cm}[x_{i},p_{j}]=i \hbar\delta
_{ij}\hspace{1.cm} [p_{i},p_{j}]=0 ,
\end{equation}
At very short scales, say string scale, the coordinates may not commute and the commutation relations are as follows :
\begin{equation}
[\hat{x}_{i},\hat{x}_{j}]=i\theta_{ij}\hspace{1.cm}[\hat{x}_{i},\hat{p}_{j}]=i\hbar\delta
_{ij}\hspace{1.cm} [\hat{p}_{i},\hat{p}_{j}]=0 ,
\end{equation}
where $\theta_{ij}$ is an antisymmetric tensor which can be defined as $\theta_{ij}=\frac{1}{2}\epsilon_{ijk}\theta_{k}$.
 The hat symbol represents the operators in the NCS.\\

\textbf{2. Hyperfine splitting in noncommutative quantum mechanics.}\\

NCQM is formulated in the same way as the standard quantum
mechanics SQM (quantum mechanics in commutative spaces), that is
in terms of the same dynamical variables represented by operators
in a Hilbert space and a state vector that evolves according to
the Schr\"{o}dinger equation :
\begin{equation}
i\hbar\frac{d}{dt}|\psi>=H_{nc}|\psi> ,
\end{equation}
 where $H_{nc}\equiv H_{\theta}$ denotes the Hamiltonian for a given
system in the NCS. It is shown in [1] that for the Hamiltonian of
the type :
\begin{equation}
H(\hat{p},\hat{x})=\frac{\hat{p}^{2}}{2m}+V(\hat{x}) .
\end{equation}
The noncommutative Hamiltonian $H_{nc}=H_{\theta}$ can be obtained
by a shift in the argument of the potential :
\begin{equation}
x_{i}=\hat{x}_{i}+\frac{1}{2}\theta_{ij}\hat{p}_{j}\hspace{2.cm}\hat{p}_{i}=p_{i} .
\end{equation}
which leads to
\begin{equation}
H_{\theta}=\frac{p^{2}}{2m}+V(x_{i}-\frac{1}{2}\theta_{ij}p_{j}) .
\end{equation}
The variables $x_{i}$ and $p_{i}$ now, satisfy in the same
commutation relations as the usual(commutative) case    i.e. equation (1). Actually equation (5)
 represents Bopp's  shift [2-4]\\
The magnetic dipole moment of the nucleus is given by :
\begin{equation}
\vec{M}=\frac{Zeg_{N}}{2M_{N}c}I
\end{equation}
where $Ze$, $M_{N}$, $I$ and $g_{N}$ are its electric charge,
mass, spin and gyromagnetic ratio respectively. In the commutative
case, the vector potential due to a point-like magnetic dipole is
given by :
\begin{equation}
\vec{A}(\vec{r})=-\frac{1}{4\pi}(\vec{M}\times \nabla )\frac{1}{r}
\end{equation}
As shown in [1], their proposal for the non commutative  H-atom
Hamiltonian can be generalized to other systems, i.e. taking the
usual Hamiltonian that is  now  a function of noncommutative
coordinates$(\hat{x},\hat{p})$, so we take the usual expression
for vector potential but now being a function of noncommutative
coordinates, accordingly this leads to a expression for the
noncommutative Hamiltonian which is the same as the usual
Hamiltonian that is  now  a function of noncommutative
coordinates, so
\begin{equation}
\hat{\vec{A}}(\vec{r})=-\frac{1}{4\pi}(\vec{M}\times \hat{\nabla} )\frac{1}{\hat{r}}
\end{equation}
where the hat symbol represents the same variable in the NCS. \\
From equation (5), we observe that $\frac{\partial}{\partial x_{i}}=\frac{\partial}{\partial \hat{x_{i}}}$, so
  $\nabla=\hat{\nabla}$. $\hat{\vec{M}}=\vec{M}$, because the noncommutativity of space has no effect on spin.\\
One can derive the magnetic field produced by the nucleus from the vector potential :
\begin{equation}
\hat{\vec{B}}=\nabla \times \hat{\vec{A}}=-\frac{\vec{M}}{4\pi}\nabla^{2}\frac{1}{\hat{r}}+\frac{1}{4\pi}\nabla (\vec{M}\cdot \vec{\nabla})\frac{1}{\hat{r}}
\end{equation}

Then the perturbative   Hamiltonian is given by :
\begin{equation}
H_{I}=-\vec{M_{e}}\cdot \vec{B}=\frac{e}{m_{e}c}\vec{S}\cdot \vec{B}=\frac{Ze^{2}g_{N}}{2m_{e}M_{N}c^{2}}\frac{1}{4\pi}
\vec{S}\cdot\left[-\vec{I}\nabla^{2}\frac{1}{\hat{r}}+\nabla(\vec{I}\cdot\vec{\nabla})\frac{1}{\hat{r}}\right]
\end{equation}
where $\vec{M_{e}}=\frac{e}{m_{e}c}\vec{S}$ is the electron magnetic dipole and $m_{e}$ is its mass.\\

  We first consider the second term in $H_{I}$}:\\
\begin{equation}
\left\langle \left((\vec{S}\cdot\vec{\nabla})(\vec{I}\cdot\vec{\nabla})\frac{1}{\hat{r}}\right)\right\rangle=
S_{i}I_{k}\left\langle \frac{\partial}{\partial\hat{x_{i}}}\frac{\partial}{\partial\hat{x_{k}}}\frac{1}{\hat{r}}\right\rangle
=S_{i}I_{k}\left\langle \frac{\partial}{\partial x_{i}}\frac{\partial}{\partial x_{k}}\frac{1}{\hat{r}}\right\rangle
\end{equation}
Using equation (5),  we have :\\

$\frac{1}{\sqrt{\hat{x}\hat{x}}}=\frac{1}{\sqrt{\hat{r}\hat{r}}}=\frac{1}{\sqrt{(x_{i}-\theta_{ij}p_{j}/2)(x_{i}-\theta_{ik}p_{k}/2)}}=$
\begin{equation}
r^{-1}\left[1+\frac{1}{r^{2}}\left(\frac{1}{2}\theta_{ij}x_{i}p_{j}-\frac{1}{8}\theta_{ij}\theta_{ik}p_{j}p_{k}+O(\theta^{3})\right)\right]
\end{equation}
which can be written as :
\begin{equation}
\frac{1}{\sqrt{\hat{r}\hat{r}}}=r^{-1}\left[1+\frac{1}{r^{2}}\left(\frac{1}{4}\vec{L}\cdot\vec{\theta}-\frac{1}{8}\theta_{ij}\theta_{ik}p_{j}p_{k}+O(\theta^{3})\right)\right]
\end{equation}

where we have used the definition $\theta_{ij}=\frac{1}{2}\epsilon_{ijk}\theta_{k}$ and $\vec{L}=\vec{r}\times \vec{p}$.
 The first term is the one in the commutative case and the other terms give the corrections due to the noncommutativity of
 space to the first and the second order.\\
One can show that  for the states with $\ell=0$ in equation (12),
the contributions of the terms with $i\neq j$ will  vanish and for
the terms  $i=j$ it  gives :
\begin{equation}
\frac{1}{3}S_{i}I_{i}\left\langle  \nabla^{2} \frac{1}{\hat{r}}\right\rangle
\end{equation}
We note that the first term in the brackets in equation (11) is $-S_{i}I_{i}\nabla^{2}\frac{1}{\hat{r}}$, so we have :\\
\begin{equation}
\left\langle H_{I}\right\rangle=-\vec{M_{e}}\cdot \vec{B}=\frac{e}{m_{e}c}\vec{S}\cdot \vec{B}=\frac{Ze^{2}g_{N}}{2m_{e}M_{N}c^{2}}\frac{1}{4\pi}
\left[-\frac{2}{3}S_{i}I_{i}\left\langle \nabla^{2}\frac{1}{\hat{r}}\right\rangle\right]
\end{equation}

Here a comment about the wave functions  is in order. It is worth
mentioning that in order to calculate $\left\langle
\frac{1}{r^{\alpha}}\right\rangle$, one should integrate over the
wave functions from $r=0$; on the other hand, the approximation we
are working at so far(dropping the terms of higher order in
$\theta$ ) is not valid for $r\leq\sqrt{\theta}$.    So for the
states with $\ell=0$, the corrections due to noncommutativity of
space on the wave functions should also take into account.
Therefore  we proceed as follows :

\begin{equation}
\left\langle \Psi+\theta\Delta^{(1)}\Psi\left|V(r)+\theta H_{I}^{(1)}+\theta^{2}H_{I}^{(2)}\right|\Psi+\theta\Delta^{(1)}\Psi\right\rangle
\end{equation}
where :
\begin{equation}
\theta H_{I}^{(1)}=-\frac{Ze^{2}g_{N}}{48\pi m_{e}M_{N}c^{2}}S_{i}I_{i}\nabla^{2}\frac{1}{r^{3}}\vec{L}.\vec{\theta}
\end{equation}
\begin{equation}
\theta^{2} H_{I}^{(2)}=\frac{Ze^{2}g_{N}}{96\pi
m_{e}M_{N}c^{2}}S_{i}I_{i}\nabla^{2}\frac{1}{r^{3}}\theta_{ij}\theta_{ik}p_{j}p_{k}
\end{equation}
 and  $V=-\frac{Ze^{2}}{r}$.\\
 We set $\theta_{3}=0$ and the rest of the $\theta$-components to zero, which can be done by a rotation
 or a redefinition of coordinates, so $\vec{L}.\vec{\theta}=L_{z}\theta$.

 Using the first-order perturbation theory for the wave functions, we have :
\begin{equation}
\Psi_{n}^{\theta}=\Psi_{n}+\sum_{k\neq n}\frac{\left\langle k\ell^{\prime}j^{\prime}j^{\prime}_{z}\left|\beta\theta L_{z} \right|n\ell j j_{z}\right\rangle}{\left|E_{n}^{0}-E_{k}^{0}\right|}\Psi_{k}
\end{equation}
where :
\begin{equation}
\beta=-\frac{Ze^{2}g_{N}}{48\pi m_{e}M_{N}c^{2}}S_{i}I_{i}\nabla^{2}\frac{1}{r^{3}}
\end{equation}
By taking into account the fact that

\begin{equation}
\left\langle \ell jj_{z}\left|L_{z}\right|\ell^{\prime}jj^{\prime}_{z}\right\rangle=j_{z}\hbar\left(1\mp \frac{1}{2\ell+1}\right)\delta_{\ell\ell^{\prime}}\delta_{j_{z}j_{z}^{\prime}},\hspace{1.cm}j=\ell\pm\frac{1}{2}
\end{equation}
 we have :

\begin{equation}
\Psi_{n}^{\theta}=\Psi_{n}+\beta\theta\sum_{k\neq n}\frac{j_{z}\hbar \left(1\mp \frac{1}{2\ell+1}\right)\delta_{\ell\ell^{\prime}}\delta_{j_{z}j_{z}^{\prime}}}{{\left|E_{n}^{0}-E_{k}^{0}\right|}}\Psi_{k}\hspace{1.cm}j=\ell\pm \frac{1}{2}
\end{equation}
The first-order correction terms are :
\begin{equation}
2\left\langle \Psi\left|V(r)\right|\theta\Delta^{(1)}\Psi \right\rangle+\left\langle \Psi\left|\theta H^{(1)}_{I}\right|\Psi \right\rangle
\end{equation}

We consider the $\ell=0$ states, so  it is obvious that  the
second term  i.e. the correction to the Hamiltonian to the first
order,  vanishes: $\theta H_{I}^{(1)}=0$.

we note that in the relation $j=\ell\pm \frac{1}{2}$, for $\ell=0$
states only the upper sign(plus) is acceptable which gives
$j=\frac{1}{2}$. But this corresponds to the upper sign(minus) in
 equation (23), which leads to $\Psi_{n}^{\theta}=\Psi_{n}$, and
this means that to the first order in $\theta$, there is no
corrections to the wave functions i.e. $\theta\Delta^{(1)}\Psi=0$.
One can show the same is true for the second-order  term
$\theta\Delta^{(2)}\Psi=0$.  Therefore,  the noncommutativity of
space to the first order has no effect on the hyperfine splitting,
so we study the noncommutativity effects to the second order. The
third term in equation (14), can be written as :
\begin{equation}
\frac{1}{8 r^{3}}(\theta_{ij}\theta_{ik})p_{j}p_{k}=\frac{1}{32 r^{3}}(\epsilon_{ij\mu}\epsilon_{ik\nu})p_{j}p_{k}\theta_{\mu}\theta_{\nu}=\frac{1}{32 r^{3}}(p^{2}\theta^{2}-(\vec{p}.\vec{\theta})^{2})
\end{equation}

where we have used the identity:
\begin{equation}
\epsilon_{ij\mu}\epsilon_{ik\nu}=\delta_{jk}\delta_{\mu\nu}-\delta_{j\nu}\delta_{k\mu}
\end{equation}
So we have :
\begin{equation}
\frac{1}{32 r^{3}}(\epsilon_{ij\mu}\epsilon_{ik\nu})p_{j}p_{k}\theta_{\mu}\theta_{\nu}=
\frac{1}{32 r^{3}}(p^{2}_{x}+p^{2}_{y})\theta^{2}
\end{equation}
and
\begin{equation}
\left\langle \nabla^{2}\left(\frac{\theta_{ij}\theta_{ik}p_{j}p_{k}}{8 r^{3}}\right)\right\rangle=
\frac{1}{8}\theta^{2}\left\langle {\left(p^{2}_{x}+p^{2}_{y}\right)}\nabla^{2}\frac{1}{r^{3}}\right\rangle
\end{equation}
By symmetry considerations we have $\left\langle p^{2}_{x}\right\rangle=\left\langle p^{2}_{y}\right\rangle=\left\langle
 p^{2}_{z}\right\rangle=\frac{1}{3}\left\langle p^{2}\right\rangle$, so the total correction due to noncommutativity of
 space on the Hamiltonian $H_{I}$ is given by :
 \begin{equation}
\left\langle
H_{I}^{(2)}\right\rangle_{\theta}=\frac{Ze^{2}g_{N}}{96\pi
m_{e}M_{N}c^{2}}\theta^{2}\vec{S}\cdot \vec{I}\left\langle
\frac{p^{2}}{r^{5}}\right\rangle
\end{equation}
To calculate the expectation value of $\frac{p^{2}}{r^{5}}$, we note that $\frac{p^{2}}{2m}+V=E$,
where $V=-\frac{Ze^{2}}{r}$, so we have :
\begin{equation}
p^{2}=2mH_{0}+2m\frac{Ze^{2}}{r}
\end{equation}
then equation (19) reads :
\begin{equation}
\theta^{2}\left\langle
H_{I}^{(2)}\right\rangle_{\theta}=\frac{Ze^{2}g_{N}}{48\pi
M_{N}c^{2}}\theta^{2}\vec{S}\cdot \vec{I}\left[E_{n}\left\langle
\frac{1}{r^{5}}\right\rangle+Ze^{2}\left\langle
\frac{1}{r^{6}}\right\rangle\right]
\end{equation}
which can be written in the following form :
\begin{equation}
\left\langle
H_{I}^{(2)}\right\rangle_{\theta}=\frac{Ze^{2}g_{N}}{48\pi
M_{N}c^{2}}\theta^{2}\vec{S}\cdot
\vec{I}\left[E_{n}f(5)+Ze^{2}f(6)\right]
\end{equation}
where $f(\alpha)=\left\langle \frac{1}{r^{\alpha}}\right\rangle$. Here $E_{n}$ is the energy eigenvalue of the
nth energy level. \\

If $F$ be the total spin of the electron and nucleus, then we have :
\begin{equation}
\frac{\vec{S}\cdot \vec{I}}{\hbar^{2}}=\frac{\vec{F}^{2}-\vec{S}^{2}-\vec{I}^{2}}{2\hbar^{2}}=\frac{\left[F(F+1)-\frac{3}{4}-I(I+1)\right]}{2}
\end{equation}
 which is $\frac{1}{2}I$ for $F=I+\frac{1}{2}$ and $\frac{1}{2}(-I-1)$ for $F=I-\frac{1}{2}$. So the total expectation
 value of the Hamiltonian $H=H_{I}+H_{\theta}$ is given by :\\

$\left\langle H_{I}\right\rangle+\left\langle H_{I}\right\rangle_{\theta}=$
\begin{equation}
\left[\frac{4}{3}\frac{mg_{n}}{M_{N}}(Z\alpha)^{4}mc^{2}\frac{1}{n^{3}}\frac{\vec{S}\cdot \vec{I}}{\hbar^{2}}+\frac{(Ze^{2})\hbar^{2}g_{N}}{48\pi M_{N}c^{2}}\theta^{2}\frac{\vec{S}\cdot \vec{I}}{\hbar^{2}}\left(-\frac{1}{2}\mu c^{2}\frac{(Z\alpha)^{2}}{n^{2}}f(5)+Ze^{2}f(6)\right)\right]
\end{equation}
The first term is the expectation value of the interaction Hamiltonian in commutative case and the rest shows
the noncommutativity effects.\\
Using $\alpha=\frac{e^{2}}{\hbar c}$, Eq.(34) gives :
\begin{equation}
\left[\frac{4}{3}\frac{mg_{n}}{M_{N}}(Z\alpha)^{4}mc^{2}\frac{1}{n^{3}}+\frac{(Ze^{2})^{2}\hbar^{2}g_{N}}{48\pi M_{N}c^{2}}\theta^{2}\left(-\frac{1}{2}\mu \frac{Ze^{2}}{n^{2}\hbar^{2}}f(5)+ f(6)\right)\right]
\end{equation}
For the  ground state   of  Hydrogen atom $n=1$, $Z=1$, and by
substituting the values of $\mu$, $e$, $\hbar$, $f(5)$ and $f(6)$
we found out that the term in the parentheses is positive and
therefore the correction due to noncommutativity of space on
$\Delta E_{NC}(F=1\rightarrow F=0)$ is positive :
\begin{equation}
\Delta E_{NC}(F=1\rightarrow F=0)=\frac{(Ze^{2})^{2}\hbar^{2}g_{N}}{48\pi M_{N}c^{2}}\theta^{2}\left(-\frac{1}{2}\mu
\frac{Z^{2}e^{2}}{n^{2}\hbar^{2}}f(5)+f(6)\right)>0
\end{equation}
This leads us to the fact that the energy difference between the
two states in a NCS is bigger than the one in commutative case, so
the radiation wavelength in $NCS_{s}$ spaces must be shorter than
the
 radiation wavelength in commutative spaces. \\
The magnitude of $\Delta E_{NC}$, for a H-atom can be obtained by
substitution of the values of various quantities as given below
:\\

$c=3\times 10^{8}ms^{-1}$, $e=1.602\times 10^{-19}C$,
$m_{e}=9.109\times 10^{-31}kg$, $M_{N}=1.673\times 10^{-27}kg$,
$\hbar=1.055\times 10^{-34}Js$, $g_{N}\simeq 5.56$  and the Bohr
radius $a_{0}=5.292\times 10^{-11}m$, which leads us to the result
: $\Delta E_{NC}=10^{-28}\theta^{2}$.\\
On the other hand one can use the data on the hyperfine splitting
to impose some bounds on the value of noncommutativity parameter
$\theta$. The best measurement of the hyperfine splitting is for
Hydrogen and its relative accuracy is about $10^{-12}$[5]. Since
the noncommutativity of space has not been detected sofar, the
value of $\Delta E_{NC}$, should be of the order of $10^{-12}$, so
we have :
\begin{equation}
\Delta E_{NC}(F=1\rightarrow
F=0)=\frac{(Ze^{2})^{2}\hbar^{2}g_{N}}{48\pi
M_{N}c^{2}}\theta^{2}\left(-\frac{1}{2}\mu
\frac{Z^{2}e^{2}}{n^{2}\hbar^{2}}f(5)+f(6)\right)\leq 10^{-12}
\end{equation}
which gives :
\begin{equation}
\theta\leq (10^{3}Gev)^{-2}
\end{equation}
This is in agreement with other results presented in the
literature, e.g.[1].\\

In conclusion, we have presented the hyperfine splitting within
the framework of  NCQM. If there exists any noncommutativity of
space in nature, as seems to emerge from different theories and
arguments, its implications should appear in hyperfine splitting 
of physical systems such as
the one treated here. We also find an upper bound for the noncommutativity parameter $\theta$. \\

\textbf{References.}\\

1. M. Chaichian, M. M. Sheikh-Jabbari and A. Tureanu,  Phys. Rev. Lett, Vol. 86, (2001), pp.2716 and
Eur.    Phys.   J.  C,  Vol.    36, (2004), pp. 251.\\
2. F. G. Scholtz et.al., Phys. Rev. D 71(2005)085005.\\
3. A. P. Balachandran et.al., hep-th/0608138.\\
4. F.G. Scholtz et.al., J. Phys. A40(2007)14581.\\
5. Essen et.al., Nature 229(1971)110.\\

\end{document}